\documentclass[aps,prl,reprint]{revtex4-2}

\usepackage{graphicx}
\usepackage{dcolumn}
\usepackage{bm}

\usepackage{pgfplots}
\usetikzlibrary{pgfplots.groupplots}
\usepackage{booktabs,caption}
\usepackage[flushleft]{threeparttable}
\usepackage{subcaption} 
\usepackage[export]{adjustbox}
\usepackage[section]{placeins}
\usepackage{remreset}
\usepackage[flushleft]{threeparttable}
\usetikzlibrary{decorations.pathreplacing}
\pgfplotsset{compat=1.16}
\usepackage{tikz}
\usepackage{tikzscale}
\usepackage{graphicx}
\usetikzlibrary{shapes}

\usepackage{xcolor}
\definecolor{tanne}{RGB}{82,121,111}	
\definecolor{lind}{RGB}{190,238,98}		
\definecolor{coral}{RGB}{242,84,91}		
\definecolor{auber}{RGB}{73,53,72}		
\definecolor{orange}{RGB}{188,50,0}	
\begin{document}

\title{Thermodynamic Description of Interfaces applying the 2PT method on ReaxFF Molecular Dynamics simulations}

\author{Christoph K. Jung}
\affiliation{Helmholtz Institute Ulm (HIU) Electrochemical Energy Storage, Helmholtzstr. 11, D-89081 Ulm, Germany}
\affiliation{Karlsruhe Institute of Technology (KIT), P.O. Box 3640, D-76021 Karlsruhe, Germany}
\author{Laura Braunwarth}
\affiliation{Institute of Electrochemistry, Ulm University, Albert-Einstein-Allee 47, D-89081 Ulm, Germany}
\author{Andrey Sinyavskiy}
\affiliation{Institute of Electrochemistry, Ulm University, Albert-Einstein-Allee 47, D-89081 Ulm, Germany}
\author{Timo Jacob} 
\email{timo.jacob@uni-ulm.de}
\affiliation{Institute of Electrochemistry, Ulm University, Albert-Einstein-Allee 47, D-89081 Ulm, Germany}
\affiliation{Helmholtz Institute Ulm (HIU) Electrochemical Energy Storage, Helmholtzstr. 11, D-89081 Ulm, Germany}
\affiliation{Karlsruhe Institute of Technology (KIT), P.O. Box 3640, D-76021 Karlsruhe, Germany}

\pagebreak

\begin{abstract}
The interface between liquid water and the Pt(111) metal surface is characterized structurally and thermodynamically via reactive molecular dynamics (MD) simulations within the ReaxFF framework. The formation of a distinct buckled adsorbate layer and subsequent wetting layers is tracked via the course of the water’s density as well as the distribution of the H\textsubscript{2}O molecules with increasing distance to the metal surface. Hereby, also the Two Phase Thermodynamics method (2PT) has been utilized for studying the course of entropy as well as the translational, rotational and vibrational entropic contributions throughout the Pt(111)\big\vert H\textsubscript{2}O interface. A significant reduction of the entropy compared to the bulk value is observed in the adsorbate layer ($S$ = 31.05$\pm$2.48\,J/molK ) along with a density of 3.26$\pm$0.06\,g/cm$^{3}$. The O-O interlayer distribution allows direct tracing of the water ordering and a quantified comparison to the ideal hexagonal adlayer. While the adsorbate layer at the Pt surface shows the occurrence of hexagonal motifs, this near-order is already weakened in the wetting layers. Bulk behavior is reached at 15\,$\mathrm{\mathring{A}}$ distance from the Pt(111) metal. Introducing an electric field of 0.1 V/$\mathrm{\mathring{A}}$ prolongs the ordering effect of the metal surface into the liquid water. 

\end{abstract}

\maketitle

\pagebreak
\section{Introduction}
In various fields of natural sciences, the investigation of a liquid's thermodynamical characteristics promises to elucidate its role in interfacial properties and processes, \textit{e.g.} ligand binding or surface chemistry. As the entropy ($S$) is one of the fundamental thermodynamic quantities, the desire to obtain estimates of the entropy directly from theoretical molecular dynamics (MD) simulations wants to be realized. While several accurate but extensive methods have been developed, their application has been restricted to small model systems.\cite{Lazaridis,Wang,Sharma,Tyka,Henchman,Karplus} Here, originally aiming to enable the study of water in different biological and chemical environments, the Two Phase Thermodynamics (2PT) method has been developed by Lin \textit{et al.},\cite{Lin03,Lin10, Pascal} yielding thermodynamic properties by merely post-processing a MD trajectory. 
The basic concept of the 2PT method is the division of the liquid system in diffusive gas-like and vibrational solid-like components. 
This is done by obtaining the density of states ($DoS$) function via a Fourier transformation of the velocity autocorrelation function (VACF) and relating the $DoS$ to a solid (same temperature and pressure) and to a hard-sphere gas (same temperature and density). Hereby, a fluidicity parameter $f$ determines the ratio of the solid and gas-like components. $f$ is described as the ratio of the (liquid system's) diffusivity to the hard-sphere gas diffusivity. 
Then, the phonon gas model allows the calculation of entropy through the vibrational density of states (\textit{e.g.} the power spectrum) by treating the system as a continuous collection of noninteracting quantum harmonic oscillators. Regarding the gas-like component, the classical theory for gases yields the thermodynamic properties. In summary, the entropy of the liquid is composed of the entropies of the solid and gas-like subsystems. This approach has already been successfully applied to study common solvents\cite{Pascal}, water\cite{Pascal12} and recently for exploring the three-dimensional environment of solvated small molecules\cite{Persson17}.\\
In this work, the 2PT method has been adapted to ReaxFF, which is a bond-order dependent reactive force field framework.\cite{Duin2001,vanDuin2} Afterwards, this approach was applied to evaluate the entropy contributions at the Pt(111)\big \vert H\textsubscript{2}O interface, which were then compared to bulk water. It is hereby the aim to elucidate the structure and dynamics of the adsorbate and wetting layers of water at room temperature. The effect of an applied electric field on the interface properties is studied as well.

\section{Two-Phase Thermodynamics method: Obtaining thermodynamic properties}
\subsection{2PT formalism}
The 2PT algorithm begins by calculating the total velocity autocorrelation function $C(t)$ as the mass weighted sum of the respective atom velocity autocorrelation functions $c_{j}^{k}$, 
\begin{eqnarray}
C(t) &&= \sum_{j=1}^{N} \sum_{k=1}^{3}  m_j c_j^k(t) \nonumber \\ &&= \sum\limits_{j = 1}^{N} \sum\limits_{k = 1}^{3} m_j \lim\limits_{\tau \to \infty} \frac{1}{2\tau}  \int_{-\tau}^{\tau} v_j^k(t'+t) v_j^k(t') dt'
\end{eqnarray}
with $m_{j}$ being the mass of the atom $j$ and $v^{k}_{j}(t)$ the $k$-th component of atom $j$'s velocity at time $t$. The total density of states $DoS(\nu)$ is calculated via a Fast Fourier Transformation of the VACF, with frequency $\nu$: 
 \begin{equation}
DoS(\nu) = \frac{2}{k_{\mathrm{B}}T} \lim\limits_{\tau \to \infty} \int_{-\tau}^{\tau} C(t) e^{-2\pi i\nu t} dt 
 \end{equation}
For polyatomic fluids, such as water, the total $DoS$ consists of translational, rotational and vibrational components. Hereby, the molecular translation is calculated via the center of mass velocities of all atoms as extracted from the ReaxFF MD trajectory, the vibrational component is based on the intramolecular vibration velocities and the rotational density is obtained via the angular velocities. 
Now, the critical point of the 2PT method is the separation of the total $DoS$ into a diffusive $DoS(\nu)_{\mathrm{gas}}$ and a solid component $DoS(\nu)_{\mathrm{solid}}$ (for a detailed motivation and the progression from the One-Phase Thermodynamics method see Refs. 7-8): 
The $DoS(\nu)_{\mathrm{solid}}$ is the density of states of a solid at the same temperature and pressure and fulfills the subsequent condition: $\lim \limits_{\nu \to 0} DoS(\nu)_{\mathrm{solid}} = 0$. The $DoS(\nu)_{\mathrm{gas}}$  is the density of states of a hard-sphere gas at the same temperature and density, including the properties $DoS(\nu)_{\mathrm{gas}}(0) = DoS(0)$ and $\lim \limits_{\nu \to \infty} DoS(\nu)_{\mathrm{gas}} = 0$.
This partition states the dynamical equivalence of the liquid system to the combination of the gas-like and solid components. Thus, the entropy $S$ of a liquid is estimated via the entropy of a hard-sphere gas and the entropy of a solid part: 
 \begin{equation}
 S = (1-f) S_{\mathrm{solid}} + f S_{\mathrm{gas}}
 \end{equation}
 Here, the fluidicity factor $f$ determines the partition ratio between the solid- and gas-like components $S_{\mathrm{solid}}$ and $S_{\mathrm{gas}}$. This correlates to a partition of the total number of particles $N$ in $fN$ hard-sphere gas particles and $(1-f)N$ solid particles, modeled by harmonic oscillators.  Thereby, the solid behavior (\textit{e.g.} $f= 0$) in the high density region and gas-like behavior in the high temperature/low density limit (\textit{e.g.} $f= 1$) is ensured. Lin \textit{et al.}\cite{Lin03} proposed the correlation of $f$ to the diffusivity, 
  \begin{equation}
  f = \frac{D(T,\rho)}{D_{0}^{\mathrm{gas}}(T,\rho)}
   \end{equation}
with $D$ being the self-diffusivity of the liquid system and $D_{0}^{\mathrm{gas}}$ the hard sphere diffusivity obtained by the Chapman-Enskog theory. The self-diffusivity can be obtained by the zero frequency intensity in the $DoS$: 
 \begin{equation}
 DoS(0) =  \frac{2}{k_{\mathrm{B}}T} \int_{-\tau}^{\tau} C(t) dt = \frac{12mND}{k_{\mathrm{B}}T}
 \end{equation} 
Now, the thermodynamics of these subsystems can be calculated by weighing the $DoS(\nu)_{\mathrm{gas}}$ and $DoS(\nu)_{\mathrm{solid}}$ with appropriate weighing functions, thereby obtaining the partition function $Q$, which is in turn linked to thermodynamic properties (\textit{e.g.} entropy $S$ among others). For this purpose, the reader may be referred to Ref. 8. 

\subsection{2PT refinement}
In 2017, Sun \textit{et al.}\cite{Sun} reanalyzed the 2PT method and proposed adaptions of some of the original formulas by Lin \textit{et al.}\cite{Lin03, Lin10} for the calculation of correct thermodynamic properties of liquids. The first observation made is the overestimation of the entropy due to a stronger decline of the $DoS$ of the hard-sphere gas than that of the actual liquid. However, in our water system (see the following chapter), even though this decline occurred at frequencies higher than 770\,cm\textsuperscript{-1} the impact on the entropy is less than 1\,\% and therefore negligible.
Second, the formula for calculating the excess entropy $S_{\mathrm{ex}}$ of hard-sphere gases as introduced by Lin \textit{et al.} needs to be modified: Originally, the entropy of the solid and hard-sphere gas subsystems are calculated as follows:
\begin{eqnarray}
  S_{\mathrm{solid}} &&= Nk_{\mathrm{B}} \int_{0}^{\infty} DoS_{\mathrm{solid}}(\nu) W_{\mathrm{solid}} d\nu \nonumber \\ \quad S_{\mathrm{gas}} &&= Nk_{\mathrm{B}}  \int_{0}^{\infty} DoS_{\mathrm{gas}}(\nu) W_{\mathrm{gas}}d\nu
\end{eqnarray}
where $W_{\mathrm{solid}}$ describes the weighing function for the solid entropy contribution and the weighing function $W_{\mathrm{gas}}$ is the sum of the contribution by the ideal gas $W_{\mathrm{IG}}$ and the excess contribution $W_{\mathrm{ex}}$: 
 \begin{equation}
 W_{\mathrm{ex}} = \frac{1}{3k_{\mathrm{B}}}S_{\mathrm{ex}}(T,\rho) = \frac{1}{3} \left[ \frac{\gamma (3\gamma-4)}{(1-\gamma)^{2}} \right]
 \end{equation}
This formula had been modified by Sun \textit{et al.} by removing a $\mathrm{ln}(z)$ term in the squared bracket, where $z$ corresponds to the compressibility of the hard-sphere gas. This removal is justified, as this $\ln(z)$ accounts for identical temperature $T$ and pressure $p$ conditions, whereas the formula for $W_{\mathrm{IG}}$ (calculating the entropy of an ideal gas) depends on the temperature $T$ and the density $\rho$. 
Third, Sun \textit{et al.} proposed a refinement of the gas-solid partition by setting $f_{\mathrm{g}}^{\delta}=D/D_{0}$, with $D$ being the diffusivity of the liquid system as obtained from $S(\nu=0)$, $D_{0}$ the diffusivity of a hard-sphere gas and $\delta$ not restricted to unity. $\delta$ is empirically chosen as physical derivations are lacking. See the Supporting Information (SI) for a detailed description. In the following analysis, the $\delta$ modification has not been included due to its empirical character.

\section{ReaxFF Methodology}
The ReaxFF reactive force field method applies a bond-order-dependent potential energy formulation in combination with a time-dependent, polarizable charge description.\cite{Duin2001,vanDuin2} The potential involves both bonding terms (\textit{e.g.} bond, angle and torsion contributions) and non-bonding interaction terms (\textit{e.g.} van der Waals, Coulomb contributions and hydrogen bonds). The bond order is updated at every iteration step depending on the local atomic environment; as such, bond formation as well as dissociation events are captured. 
The partial charge of each atom is calculated by the self-consistent electron equilibration method (EEM) developed by Mortier \textit{et al.}\cite{Mortier1986}, thereby describing the electrostatic interactions. Our self-developed Pt/O/H reactive force field by D. Fantauzzi \textit{et al.}\cite{DF} has been used throughout the present work. All ReaxFF calculations were carried out in the ADF software package (version 2019.103).\cite{teVelde2001,ADF2} 
During the reactive molecular dynamics simulations, a 0.25\,fs timestep was employed to integrate the equations of motion by utilizing a velocity-Verlet algorithm. The temperature of 298.15\,K has been controlled via a Nosé-Hoover thermostat using a damping constant of 100\,fs.\cite{Nose,Hoover} 
As no ions were included in the simulations, the conditions at the clean Pt(111)\big \vert H\textsubscript{2}O interface correspond to the potential of zero charge. 
Further, in order to mimic the electrostatic potential within the interface, an additional external electric field was applied normal to the surface plane, where the field strength was varied between 0.01 and 0.25\,V/$\mathrm{\mathring{A}}$. This electric field leads to an additional acceleration on all atoms depending on their respective atomic charge. In the following, results are presented exemplary for a field strength of 0.1\,V/$\mathrm{\mathring{A}}$. See the Supporting Information (SI) for further information. 
The bulk water was simulated by 3040 water molecules in a simulation box of 45\,$\mathrm{\mathring{A}}$ edge length, corresponding to a density of $\rho = $ 0.997\,g$/$cm\textsuperscript{3}. Simulations of the Pt(111)\big \vert H\textsubscript{2}O interface systems were performed on a symmetric twelve-layer slab ($\mathrm{8 \times 8 \times 12}$ atoms), in which the central two layers were fixed to the corresponding calculated bulk crystal structure, surrounded by 60\,$\mathrm{\mathring{A}}$ of water ($\rho=$ 0.997\,g$/$cm\textsuperscript{3}) on both sides, \textit{e.g.} 1728 water molecules.
For equilibration, a MD simulation in the canonical ensemble ($NVT$) has been performed for 400,000 iterations, followed by 50,000 iterations in a microcanonical ensemble ($NVE$) and lastly 5,000 iterations (again $NVE$) have been used for the 2PT analyses. For each system, 10 independent simulations have been performed to capture statistical variations. \\

\section{Results and Discussion}
\subsection{Water at the Pt(111) interface}
\subsubsection{Bulk-H\textsubscript{2}O characterization}
\begin{figure*} 
\centering 
\includegraphics[scale=0.8]{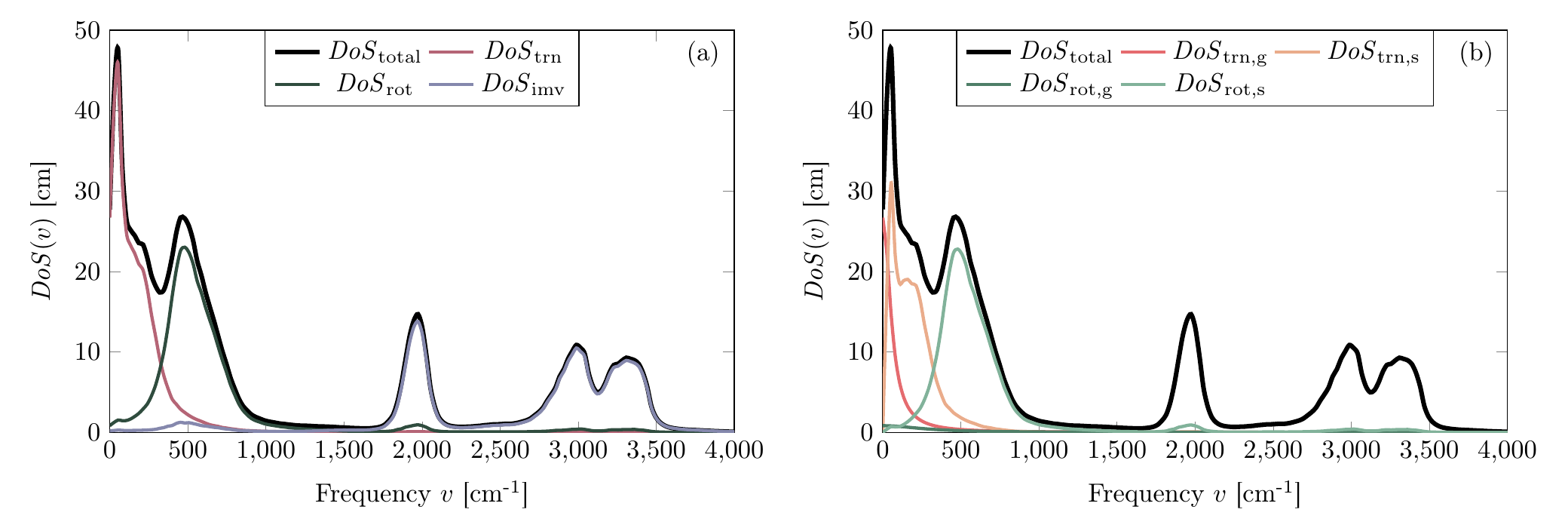}
\caption{Partition of the density of states function of liquid water in the 2PT model. (a) Total density of states function $DoS$\textsubscript{total} and the translational ($DoS$\textsubscript{trn}), rotational ($DoS$\textsubscript{rot}) and vibrational ($DoS$\textsubscript{imv}) contributions. (b) $DoS$\textsubscript{total} is obtained as a superposition of gas-like components ($DoS$\textsubscript{$\ast,$g}) and solid components ($DoS$\textsubscript{$\ast$,s}).}
\label{split}
\end{figure*}

In the first step we investigated the entropy distribution in bulk water. Figure \ref{split} shows the calculated $DoS$ for liquid water: The characteristic contributions of the three degrees of freedom for water to the total $DoS$ are depicted in (a). The first peak at 45\,cm\textsuperscript{-1} with the shoulder at 200\,cm\textsuperscript{-1} corresponds to translational librations, whereas the next peak at 514\,cm\textsuperscript{-1} is caused by rotational librations of water molecules. In (b), the gas-like component of the $DoS$ takes its maximum at $v$ = 0\,cm\textsuperscript{-1} and decays for increasing frequency while the solid contribution is dominant for higher frequency values. From the density of state functions, the total entropy of the system as well as the respective degree of freedom's contributions can be determined.

Using ReaxFF in conjunction with the 2PT method for the evaluation of a liquid water box yielded the reference values for entropy and density of bulk H\textsubscript{2}O. As can be taken from Fig. \ref{interface} (dashed blue line), we obtained a mean value over multiple independent simulations of $S_{\mathrm{bulk-H_{2}O}}$=59.27$\pm$0.52\,J/molK. Compared to the experimental value of $S_{\mathrm{exp}}$=69.95\,J/molK\cite{ExpS}, a deviation of 15\% can be observed. However, this underestimation of entropy occurs over a range of different water models, where depending on the forcefield variations in the range of -6 to -27\% are obtained (see Ref. [10]), with TIP3P performing best (deviation of +4\%), see the Supporting Information (SI). The main factor for this underestimation of entropy has been identified by Pascal \textit{et al.} as too stiff hydrogen bonding interactions leading to restricted low rattling motion.\cite{Pascal12} Thereby, ordering is enforced in liquid water, lowering the calculated entropy value. Another factor is the underrated diffusivity of water with our force field, thereby the translational entropy is undervalued and in consequence the total entropy $S$. Regarding the contributions of the three degrees of freedom to the water's entropy, the vibrational part is negligible with $\leq$0.1\,\%, rotations account for 18.6\,\% and translational motions for 81.3\,\%. See the SI for a detailed listing of the aforementioned observations. 

\subsubsection{Pt(111)\big \vert H\textsubscript{2}O interface}
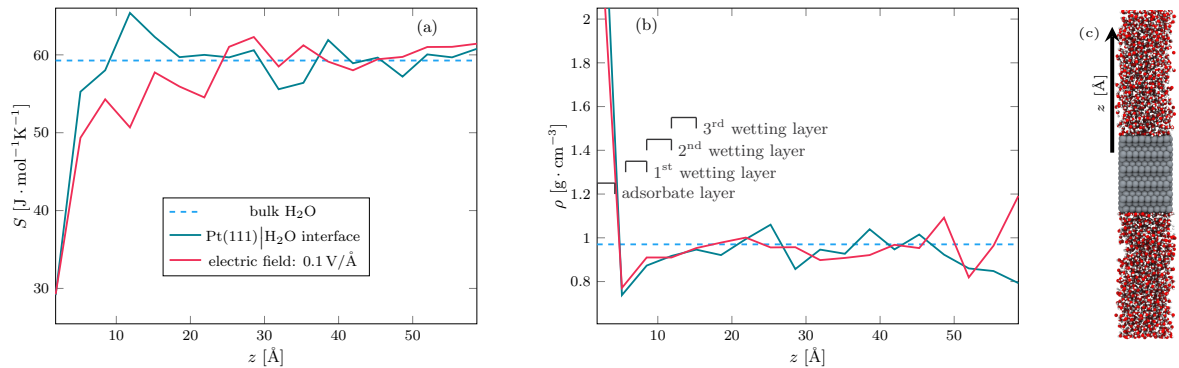
\begin{figure*} 
\centering 
	\begin{minipage}[t]{0.4\textwidth}
	\centering 
	\vspace{0pt}
	\includegraphics[scale=0.7]{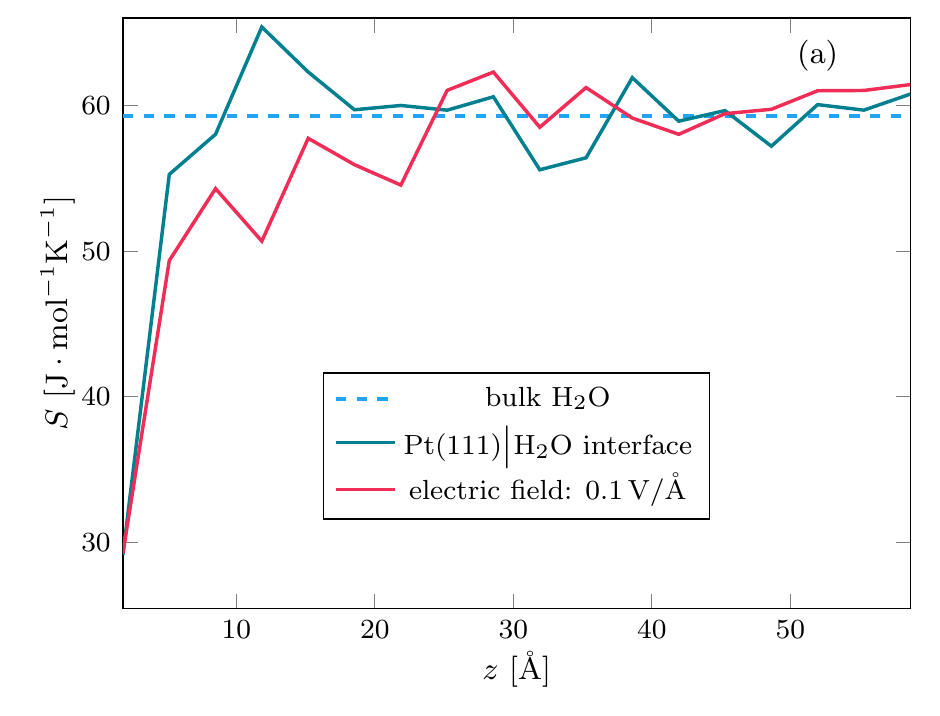}
	\end{minipage}
	\begin{minipage}[t]{0.4\textwidth}
	\centering 
	\vspace{0pt}
	\includegraphics[scale=0.7]{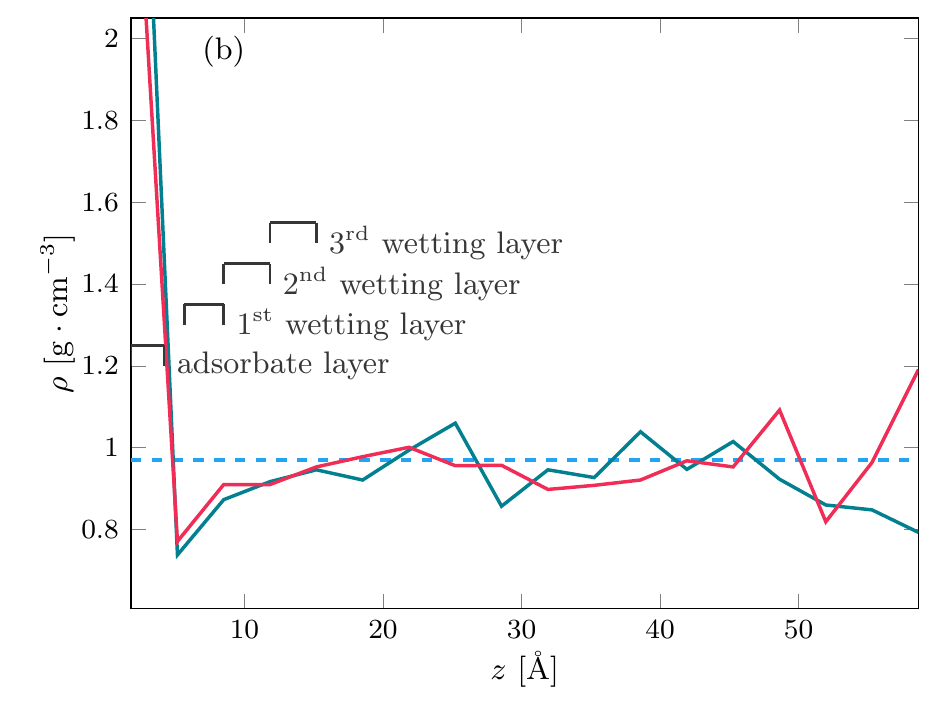}
	\end{minipage}
	\adjustbox{valign=t}{ 
	\begin{minipage}[t]{0.1\textwidth}
	\centering 
	\vspace{12pt}
	\begin{tikzpicture}
	\node[anchor=south west,inner sep=0] (Bild) at (0,0) {\includegraphics[scale=0.085]{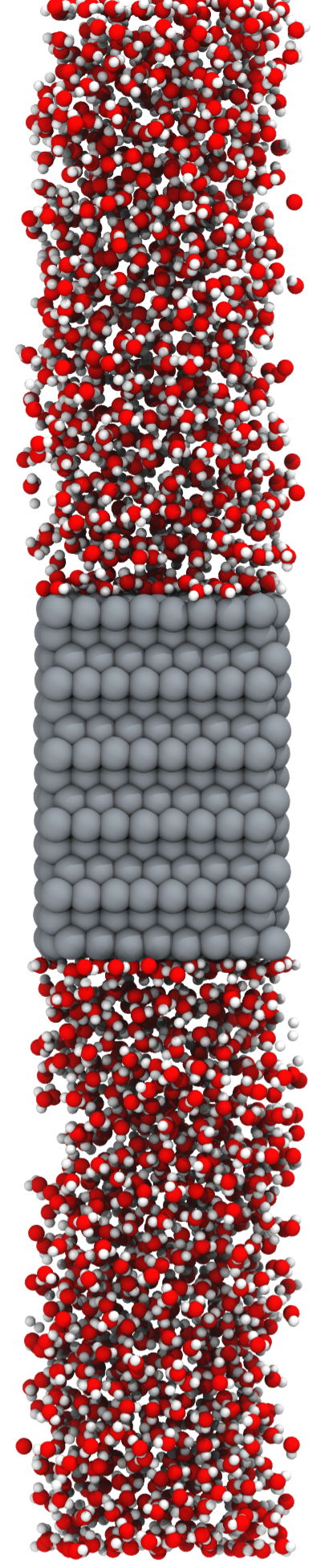}};
		\begin{scope}[x=(Bild.south east),y=(Bild.north west)]
        			\draw[black, thick, ->, line width = 1.5pt, >=stealth] (0.02,0.57) -- (0.02,0.95) ;
			\node[rotate=90] (A) at (-0.15,0.75) {\tiny $z$ [\AA]};
			\node at (-0.3, 0.93) {\tiny (c)};
    		\end{scope}
	\end{tikzpicture}
	\end{minipage}}
	\caption{(a), (b): Curves of the entropy and density at the Pt(111)\big\vert H\textsubscript{2}O interface as extracted from the 2PT method. The green (red) line shows the course of $S$ and $\rho$ with increasing distance from the Pt surface (and an applied electric field normal to the electrode of 0.1\,V/$\mathrm{\mathring{A}}$). Each data point has been obtained in a layer of $\Delta z =$ 3.35\,$\mathrm{\mathring{A}}$. 
The dashed blue lines represent the averaged mean values of $S$ and $\rho$. (c): Schematic model of the periodically continued Pt(111)\big \vert H\textsubscript{2}O system is displayed.}
\label{interface}
\end{figure*}
\begin{figure}[b] 
\centering 
\includegraphics[width = 0.49\textwidth]{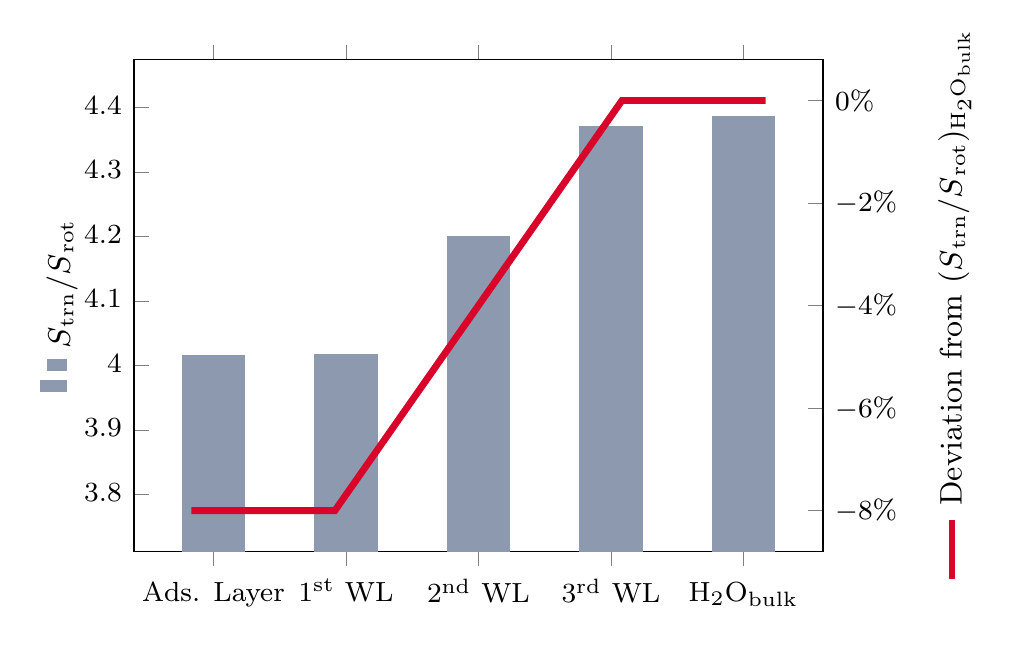}
\caption{Ratio of the translational ($S$\textsubscript{trn}) and rotational ($S$\textsubscript{rot}) contribution to the entropy with increasing distance from the platinum surface. The adsorbate layer is followed by the three subsequent wetting layers, denoted WL. For comparison, the deviation from the bulk water's $S$\textsubscript{trn}/$S$\textsubscript{rot} ratio is given in percent on the right axis.}
\label{transrot}
\end{figure}
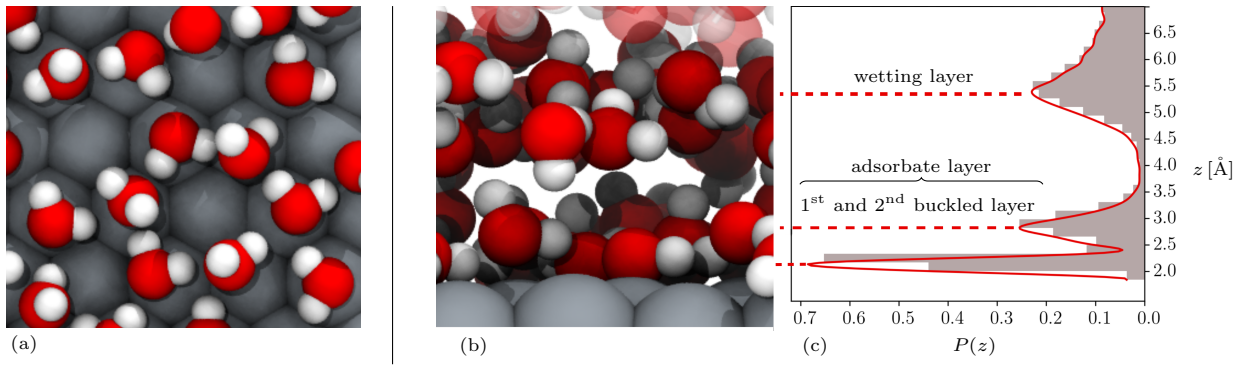
\begin{figure*}
\centering 
	\adjustbox{valign=t}{
	\begin{minipage}{0.3\textwidth}
	\centering 
		\begin{tikzpicture}
			\node[anchor=south west,inner sep=0] (Bild) at (0,0) {\includegraphics[scale=0.61]{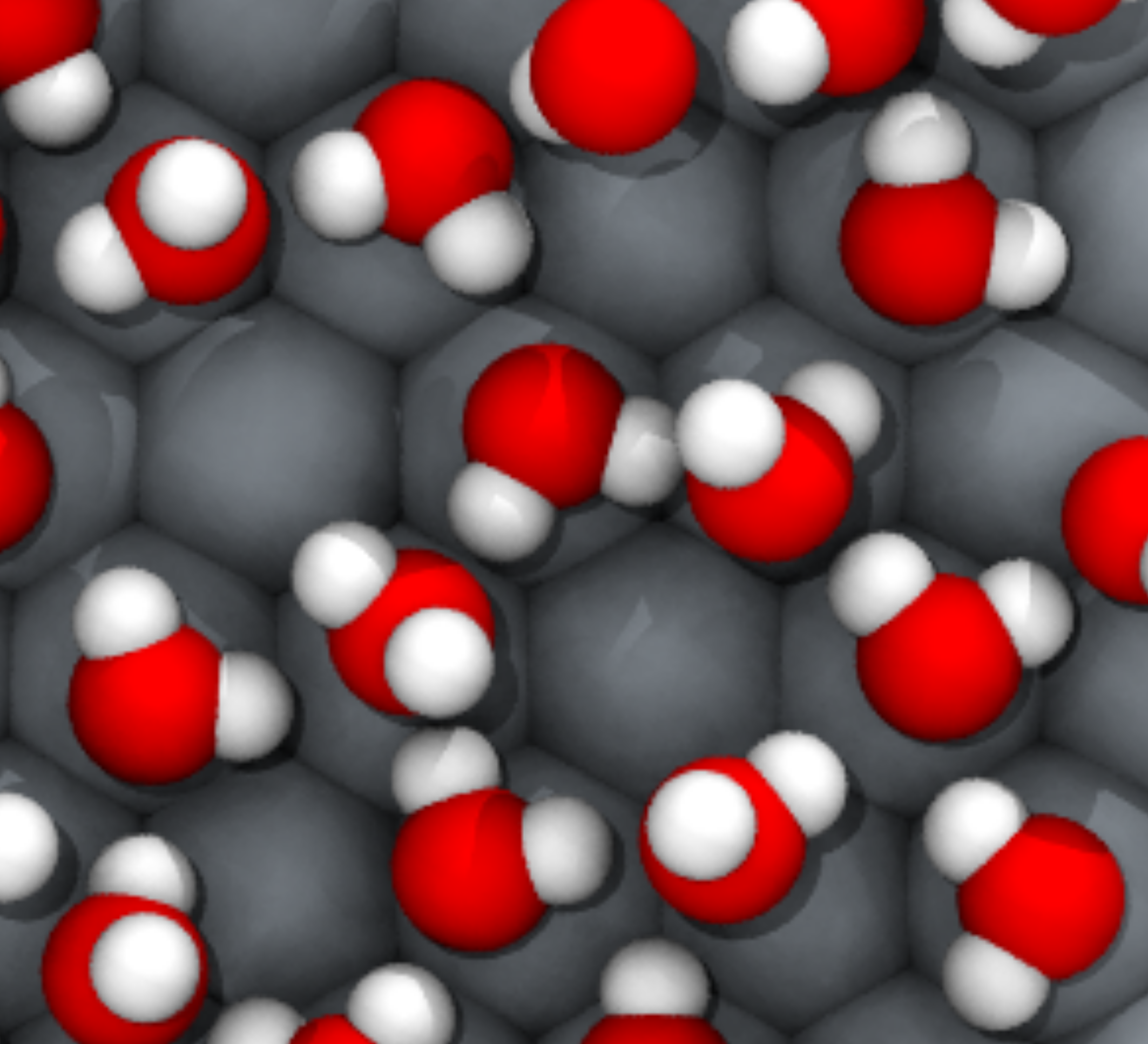}};
			\begin{scope}[x=(Bild.south east),y=(Bild.north west)]
			\node (aa) at (0.05,-0.05) {\scriptsize (a)};
			\end{scope}
		\end{tikzpicture}
	\end{minipage}}
\vline
	\adjustbox{valign=t}{ 
	\begin{minipage}{0.65\textwidth}
	\centering 
	\begin{tikzpicture}
	\node[anchor=south west,inner sep=0] (Bild) at (0,0) {\includegraphics[scale=0.5]{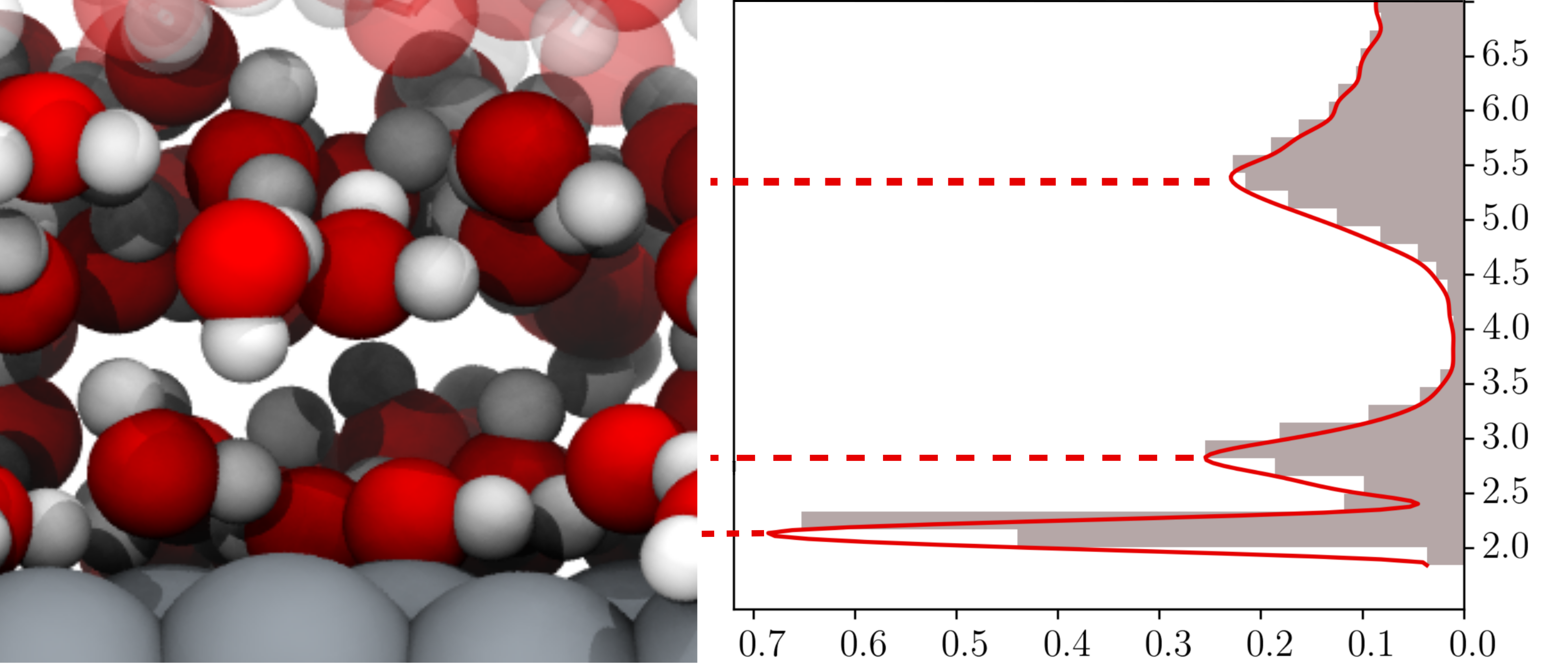}};
		\begin{scope}[x=(Bild.south east),y=(Bild.north west)]
			\node (A) at (0.71,-0.05) {\scriptsize $P(z)$};
			\node (B) at (1.025,0.5) {\scriptsize $z$\,[$\mathrm{\mathring{A}}$]};
			\node (C) at (0.635,0.39) {\scriptsize 1\textsuperscript{st} and 2\textsuperscript{nd} buckled layer};
			 \draw[-,decorate,decoration=brace] (0.48,0.45)  -- node[above] {\scriptsize adsorbate layer} (0.8,0.45);
			\node (D) at (0.63,0.78) {\scriptsize wetting layer};
			\node (bb) at (0.05,-0.05) {\scriptsize (b)};
			\node (cc) at (0.5,-0.05) {\scriptsize (c)};
    		\end{scope}
	\end{tikzpicture}
	\end{minipage}}
	\caption{(a) Top view on the buckled layers of water molecules on the Pt(111) surface (Pt: gray, O: red, H: white). (b) Side view at the Pt(111)\big\vert H\textsubscript{2}O interface. (c) Plot of the oxygen distribution in $z$ direction with increasing distance from the Pt surface. The histogram has been computed from the MD simulation used for the 2PT method and averaged over 10 independent simulations. The red curve is calculated as spline interpolation. }
\label{surf}
\end{figure*}

As can be seen in Fig. \ref{interface} (gray line), the entropy of the water molecules is significantly reduced near the Pt surface ($S$ = 31.05$\pm$2.48\,J/molK) along with an increased density ($\rho$=3.26$\pm$0.06\,g/cm$^{3}$), regaining the respective bulk values at approximately $15-20$\,$\mathrm{\mathring{A}}$ distance from the surface. Although obvious, it should be noted for any surface-water interface calculation: To allow for a formation of the complete interface system (\textit{e.g.} the adsorbate water layer, the surface-water interface and the water bulk) one should model the system with at least $15-20$\,$\mathrm{\mathring{A}}$ of water in the respective spatial directions. \\
Applying an external electric field of 0.1\,V/$\mathrm{\mathring{A}}$ amplifies the so far observed effects: The red curve in the entropy plot of Fig. \ref{interface} expresses a slower increase towards the bulk water value $S_{\mathrm{H_{2}O}}$, while the density is seemingly uninfluenced by the electric field (the red and green density curve show a similar behavior in Fig. \ref{interface}). This allows for the hypothesis that the interface character is extended and the hexagonal-like ordering of the water molecules continues further in the water bulk under the applied electric field. \\
From now on, the following naming scheme is utilized: The H\textsubscript{2}O adsorbed on the Pt(111) surface including the buckled (\textit{e.g.} lifted in $z$-direction) water molecules are denoted adsorbate layer. This layer is characterized by a significantly lowered entropy and an increased density (see Fig. \ref{interface}). The subsequent layers on top of the adsorbate layer are denoted wetting layers. There, the entropy and density curves are converting to their respective bulk behavior, however, the influence of the platinum surface is still visible. \\
From the density of states function calculated exclusively for the water molecules in the adsorbate layer, the contributions of translation, rotation and vibration can be determined. As the total entropy near the platinum surface is reduced nearly by half compared to liquid H\textsubscript{2}O, the ratio of the translational and rotational entropy contributions is also altered: As can be seen in figure \ref{transrot}, this ratio $S_{\mathrm{trn}}/S_{\mathrm{rot}}$ is reduced from 4.4 in liquid H\textsubscript{2}O to 4 in the adsorbate and first wetting layer. This corresponds to a decrease of 8\,\%. From the total entropy of H\textsubscript{2}O in the adsorbate layer account 79.9\,\% for translational and 19.9\,\% for rotational contributions. Though, we focus on the ratio instead of absolute values in favor of an increased significance, as it can also be correlated to the hydrogen bonding strength (as mentioned above). A tentative explanation for the ratio's decrease can be found from the visualization of the Pt\big\vert water interface in Fig. \ref{surf} (b): The depleted region between the adsorbate and first wetting layer may facilitate rotation, with lesser influence on the molecular translation. The structural characteristics of these regions are further discussed in the following sections. 

During the MD simulation used for the 2PT evaluation (5,000 iterations [1.25\,ps]) we assume it to be unlikely that interlayer particle exchange has taken place. The mean path length during this time interval was calculated to be $\sim$0.4\,$\mathrm{\mathring{A}}$, supporting this assumption. Tracking the movement in $z$- resp. $x$,$y$-direction of the water molecules of the layers chosen for the 2PT evaluation for longer times (\textit{e.g.} 25\,ps) leads to the following observations: Adsorbed H\textsubscript{2}O molecules remain mainly stationary. Water molecules belonging to the wetting layers express still restricted movement in $x$ and $y$-direction, though layer exchange into the adsorbate layer or subsequent wetting layers is possible. With increasing distance from the Pt(111) surface, the movement gets random and interlayer exchange is observed frequently. \\
\paragraph{Adsorbate layer}
Coming back to the reduced entropy and increased density of water near the Pt surface: It is widely known both theoretically and experimentally, that water forms a bilayer-like structure on a Pt(111) surface.\cite{Schnur2009,Gross,Langmuir,Lang2,Lang3,Lang4}. Hereby, the water molecules show a mostly hexagonal ordering in so-called H-up and H-down configurations, possibly including ``575757'' defects in the water lattice.\cite{Lang2} These defects are formed when one rotated H\textsubscript{2}O hexamer replaces a linkage in the formerly ordered hexagonal network, thereby creating new pentagons and heptagons. The H\textsubscript{2}O-coverage ($\theta_{\mathrm{H_{2}O}}$) has been established to be 0.67-0.72\,ML, depending on the presence of the mentioned defects.\cite{Langmuir,TJjacs} In our simulations, a mean coverage of $\theta_{\mathrm{H_{2}O}}$=0.64\,ML has been observed. It can be seen in the left image of Fig. \ref{surf} how the H\textsubscript{2}O molecules form five-, six- or seven-membered rings. Also, the observed entropy value of approximately 31.05$\pm$2.48\,J/molK can now directly be connected to an increased ordering in the adsorbate layer. However, considering the dynamic nature of our simulations in accordance with the applied temperature of 298\,K it is reasonable that no completely ordered ideal network can be observed.\cite{Schnur2009} Motifs like the hexagonal shapes are detectable and are consequently related to perfect bilayer structures, yet unordered, chaotic areas are also present. \\
Taking a closer look at the Pt(111)\big\vert H\textsubscript{2}O contact region, one can distinguish between a first and second buckled layer in the adsorbate layer. From the sideview image in Fig. \ref{surf} (b) the difference in oxygen height of these two layers can be observed, rising from the competition between Pt-H\textsubscript{2}O interaction and hydrogen bonding between water molecules.\cite{Langmuir} Comparing to the oxygen distribution plot in Fig. \ref{surf} (c), the first buckled layer is $\sim$2.2\,$\mathrm{\mathring{A}}$ away from the Pt surface and the second buckled layer is at $\sim$2.9\,$\mathrm{\mathring{A}}$. This is relatable to the results regarding water bilayer structure geometry obtained by density functional theory (DFT): The H-up/H-down structures are composed by half of H\textsubscript{2}O molecules laying parallel to the Pt(111) surface at a distance of 2.5\,$\mathrm{\mathring{A}}$ and the other half by H\textsubscript{2}O molecules, where one hydrogen atom is either pointing away (\textit{e.g.} H-up) or towards (\textit{e.g.} H-down) the surface.\cite{TJjacs} Hereby, the H-up or H-down water molecules are at a distance of $2.9-3.0$\,$\mathrm{\mathring{A}}$, the O\textsubscript{up/down}$-$O\textsubscript{parallel} vertical distance is in both cases 0.42\,$\mathrm{\mathring{A}}$.\cite{TJjacs} In our simulations, the observed smaller distance of $\sim$2.2\,$\mathrm{\mathring{A}}$ of the water molecules to the Pt surface is inherent to the used ReaxFF force field: Structural optimization of a single adsorbed parallel water molecule yields a distance of 2.18\,$\mathrm{\mathring{A}}$. The distance to the second buckled layer is $-$ considering the thermal motion of the water molecules $-$ comparable to the structures obtained from ideal structures in DFT. 
Subsequently after the buckled layers, a short range of $\sim$2\,$\mathrm{\mathring{A}}$ nearly depleted of H\textsubscript{2}O can be observed. This corresponds to the area of low density, namely 0.81$\pm$0.05\,gcm\textsuperscript{-3} in Fig. \ref{interface}. This unpopulated region is in accordance with simulations of multilayer water films by Antony \textit{et al.}: They observed the depleted region in $3.1-4.7$\,$\mathrm{\mathring{A}}$ distance from the surface.\cite{Langmuir} Therein, dangling hydrogen atoms can be seen pointing towards oxygen atoms beyond the depleted region enabling hydrogen bonding. 

\paragraph{Wetting layers}
The layers subsequent to the depleted region in the Pt(111)\big \vert H\textsubscript{2}O interface are denoted first, second etc.\ wetting layers. 
To investigate the ordering of water molecules depending on the distance from the Pt(111) surface, we calculated the normalized distributions of O$-$O-distances (in $x$ and $y$ direction) for the water layers. By comparing the distribution to an ideal hexagonal water network structure, the degree of ordering could be estimated as a function of $z$ and compared with and without applied electric field.

\begin{figure*} 
\centering 
\includegraphics[scale=0.67]{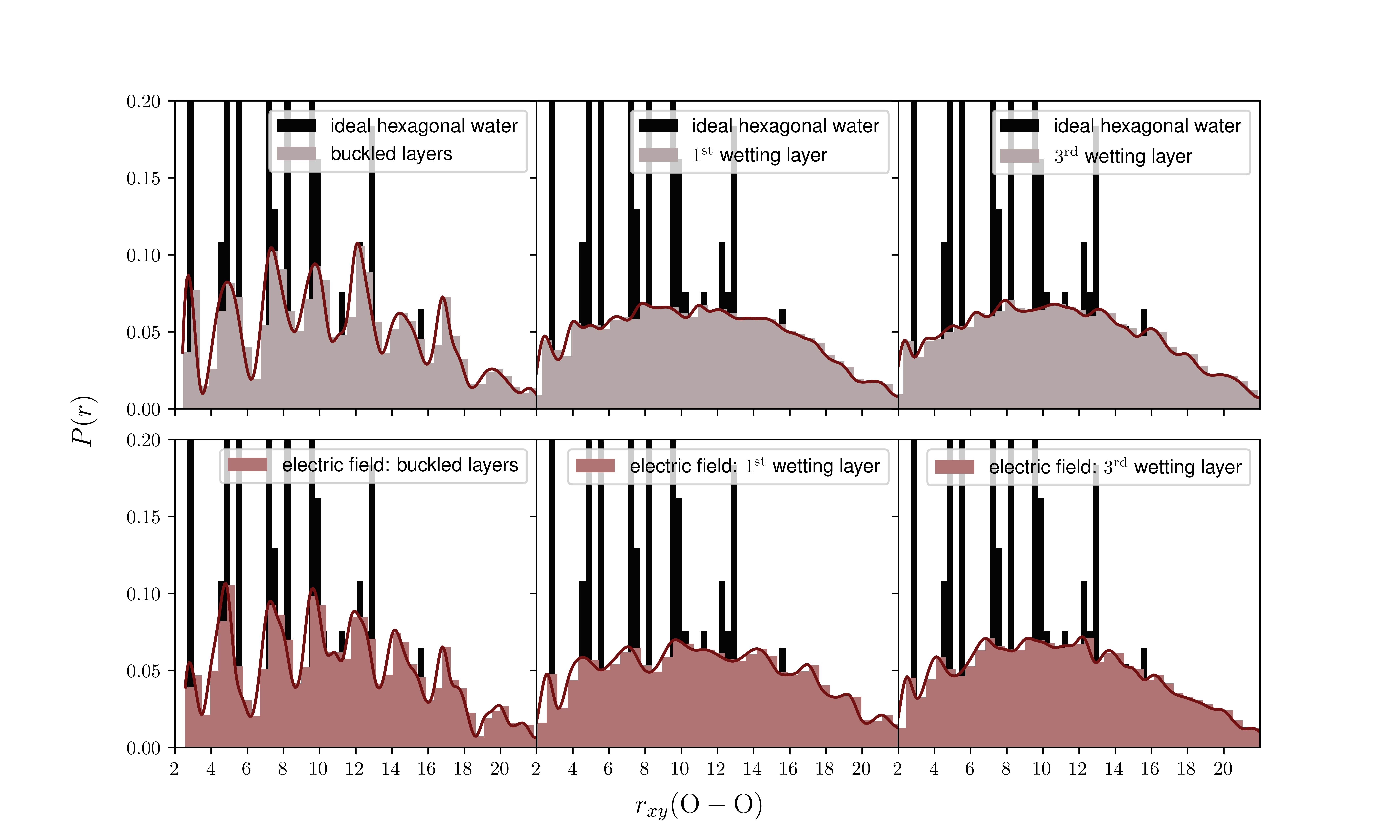}
\caption{Oxygen-oxygen distribution histograms calculated from the in-plane distance ($x$,$y$ direction). The black bars correspond to O$-$O distances calculated from an ideal hexagonal water adsorbate network. The gray (resp. light red) bars have been obtained from and averaged over 10 independent 2PT-MD-simulations (resp. applying an external electric field of 0.1\,V/$\mathrm{\mathring{A}}$). For allocation of the layer denotations see Fig. \ref{interface} (c). The red curve is obtained as spline interpolation. }
\label{roo}
\end{figure*}

In Fig. \ref{roo}, the distributions of in-plane O$-$O distances are depicted: In the buckled layers (\textit{e.g.} H\textsubscript{2}O adsorbate layer on the surface) the peaks correspond to discrete distances between the oxygens, which compares well to the ideal hexagonal network O$-$O distances. There, the degree of ordering is comparable with and without applied electric fields, aligning with the similar entropy values calculated for this layer. In the first wetting layer, corresponding to a distance range of $\sim$\,5.1$-$8.5\,$\mathrm{\mathring{A}}$ from the Pt surface, a lower entropy value can be observed under the influence of an applied electric field (see Fig. \ref{interface}). This is reflected in the histograms in Fig. \ref{roo}. Without electric field, the peaks are -- though less discrete -- clearly visible and are still shaping the overall form of the curve. At a distance range of $\sim$\,11.8$-$15.2\,$\mathrm{\mathring{A}}$ from the Pt surface (\textit{e.g.} third wetting layer), the entropy value of the water without electric field is already near its bulk value, while the application of an electric field still causes entropy values being 18\% below the bulk value. In Fig. \ref{roo}, this suggested remaining ordering of the third wetting layer is still recognizable by distinguishable peaks in the O$-$O distribution. Without the electric field, the distribution of O$-$O distances is smooth and suggests an ``chaotic'' water network, one would suggest for liquid bulk water. Thereafter, the oxygen distribution, the entropy and the density start expressing their respective bulk behavior. \\
The course of the entropy in water as obtained from 2PT calculations in conjunction with the presented in-plane O$-$O-distance distribution is therefore a valid indicator to characterize the Pt(111)\big\vert H\textsubscript{2}O interface. Applying an external electric field prolongs the interface character (\textit{i.e.} the effect of the surface on the ordering of the water network) with increasing distance from the surface. 

\FloatBarrier

\section*{Conclusion}
To summarize, we have detailed the implementation of the 2PT-method within the ReaxFF framework, taking the Pt(111)\big \vert H\textsubscript{2}O interface at the conditions of a potential of zero charge as exemplary application. We observed a significantly lowered entropy value of the adsorbate water layer at the Pt surface, concomitant with an increased density. Here, we dissolved the translational, rotational and vibrational contributions to the entropy and observed an increased ratio of translational to rotational entropic contributions compared to liquid water. This states a further conformation of the experimentally and theoretically predicted ordering of water molecules in contact with the metal surface. The density and entropy values reach their bulk limit at a distance of 15\,$\mathrm{\mathring{A}}$, therefore giving an orientation for the minimum height of the water region when modeling water as solvent or electrolyte. By analyzing the interlayer oxygen$-$oxygen-distribution, we were able to identify the buckled adsorbate layer as well as subsequent wetting layers and the ordering within. In the adsorbate layer hexagonal motifs could be detected, depicting the ordered character and reasoning the low entropy value of $S$ = 31.05$\pm$2.48\,J/molK.  This ordering is preserved better when applying an external electric field of 0.1\,V/$\mathrm{\mathring{A}}$. The 2PT method in combination with ReaxFF proves a valuable tool for studying thermodynamic properties and convinces further with its transferability and rapidly converging calculation. In future work, the influence of the electrode potential will be mimicked by setting the corresponding surface composition of hydrogen adsorbates or oxygenated intermediates and the effects on the structure and thermodynamics of the Pt(111)\big \vert H\textsubscript{2}O interface will be investigated. 

\section*{Acknowledgement}
The authors acknowledge support by the DFG (German Science Foundation) within the framework of the Collaborative Research Centers SFB-1316 as well as SFB-1249. In addition, support from the BMBF (Bundesministerium f\"ur Bildung und Forschung) through the project GEP (``Fundamentals of electrochemical interface'', Grant Agreement: 13XP5023D) is gratefully acknowledged. Further, the authors acknowledge the computer time supported by the state of Baden-W\"urttemberg through the bwHPC project and the DFG through grant number INST40/ 467-1 FUGG. \\
\clearpage

\pagebreak
\section*{Supporting Information}
As pointed out by Sun \textit{et al.}\cite{Sun}, the introduction of the $\delta$ in $f_{\mathrm{g}}^{\delta}=D/D_{0}$ dictates the partition of gas-solid components. $f_{\mathrm{g}}$ in general is called the gas fraction and should be 0 when the diffusivity of the system is 0 and approaching 1 when nearing the low density-high temperature limit.\cite{Sun} However, the authors stated this assumption formulated by Lin \textit{et al.}\cite{Lin03,Lin10} is not explicitly physically required. Including the exponent $\delta$ allows the determination of the optimal $f_{\mathrm{g}}$, depending on the respective system. Concomitant, the exact entropy can be computed. The drawback of the $\delta$-introduction is the lack of physical constraints for its determination, requiring an empirical definition of $\delta$. Exemplary, see the influence of $\delta$ for the entropy calculation of water. In our application for H\textsubscript{2}O, $\delta=1.84$ yields the best agreement to the experimental entropy of water: $S_{\mathrm{\delta, H_{2}O}}$=67.57\,J/molK, thereby shrinking the deviation to 3\% ($S_{\mathrm{exp}}$=69.95\,J/molK). \\

\begin{figure} [h!]
\centering 
\includegraphics[width = 0.49\textwidth]{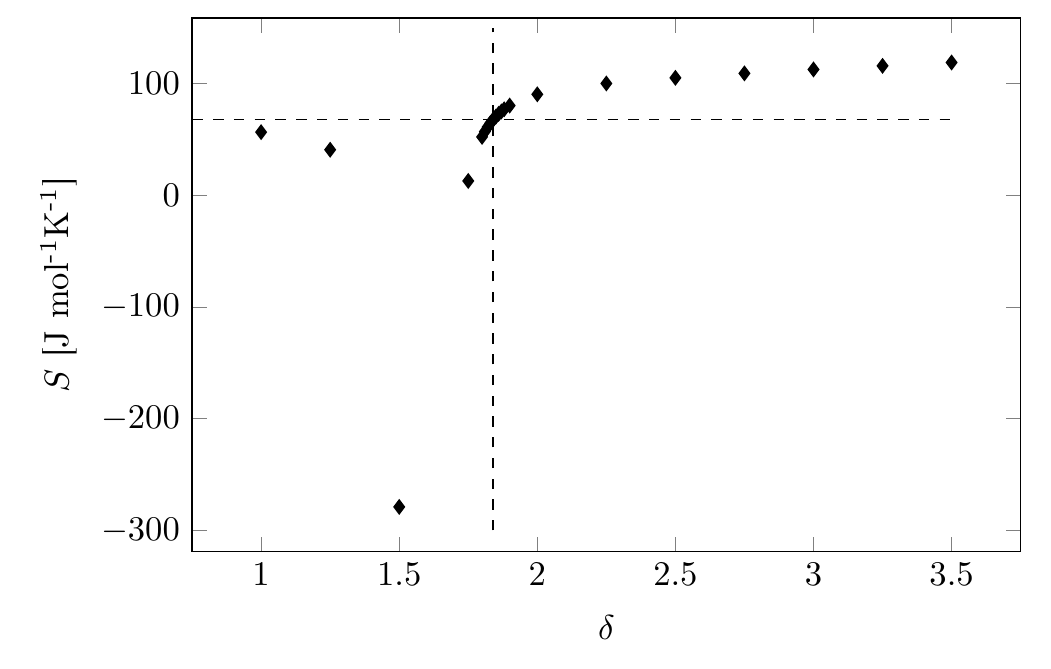}
\caption{Influence of the variation of $\delta$ in $f_{\mathrm{g}}^{\delta}=D/D_{0}$ on the 2PT calculated entropy $S$. Dashed lines denote the optimal entropy of 67.57 J/molK and the respective value for $\delta$=1.84. }
\label{vs}
\end{figure}

\ \\
For discussing the entropy value obtained by our ReaxFF simulations (computational details in section ``ReaxFF methodology''), the ratio of the translational ($S_{\mathrm{trans}}$) to the total entropy ($S_{\mathrm{tot}}$), the ratio of $S_{\mathrm{trans}}$ to the rotational entropy contribution ($S_{\mathrm{rot}}$) as well as the self-diffusivity constant of H\textsubscript{2}O ($D$) are denoted in table \ref{123}. 
\begin{table}[b]
\caption{The ratio of the translational ($S_{\mathrm{trans}}$) to the total entropy ($S_{\mathrm{tot}}$), the ratio of $S_{\mathrm{trans}}$ to the rotational entropy contribution ($S_{\mathrm{rot}}$) and the self-diffusivity constant of H\textsubscript{2}O ($D$) for different force fields.  \label{123}}
\begin{center}
  \begin{threeparttable}
	\begin{tabular}{c|ccc|c} 
       		\toprule
  & $S_{\mathrm{tot}}$ [J/molK] & $S_{\mathrm{trans}}/S_{\mathrm{tot}}$ &$ S_{\mathrm{trans}}/S_{\mathrm{rot}}$ & $D$ [$\times$ 10\textsuperscript{5} cm\textsuperscript{2}/s] \\ 
		\midrule
Exp. 		& 	69.95 			&  		&  		 & 2.27$^{a}$ \\
ReaxFF 	& 	59.27 $\pm$ 0.52 	& 0.81 	& 4.28 	& 1.36  \\
TIP3P$^{a}$	&	72.51 $\pm$ 0.27	& 0.80	& 4.05	& 5.69 \\
TIP4P$^{a}$ &	52.41 $\pm$ 0.27	& 0.85	& 5.75  	& 1.25 \\
		\bottomrule
	\end{tabular}
\begin{tablenotes}
		\small
		\item $ ^{a}$ The data has been adopted from Ref. 10. 
	\end{tablenotes}
\end{threeparttable}
\end{center}
\end{table}
From the ratio $S_{\mathrm{trans}}$/$S_{\mathrm{rot}}$ one can infer enhanced hydrogen bonding strength, thereby inducing a stronger ordering behavior in liquid water resulting in a lower $S_{\mathrm{tot}}$ value. This is also mirrored in the self-diffusivity constant obtained in our ReaxFF calculations: It is ca. 4 times lower than TIP3P's self-diffusivity corresponding to stiffer hydrogen bonding network. \\ \

\begin{figure} [h!]
\centering 
\includegraphics[width = 0.49\textwidth]{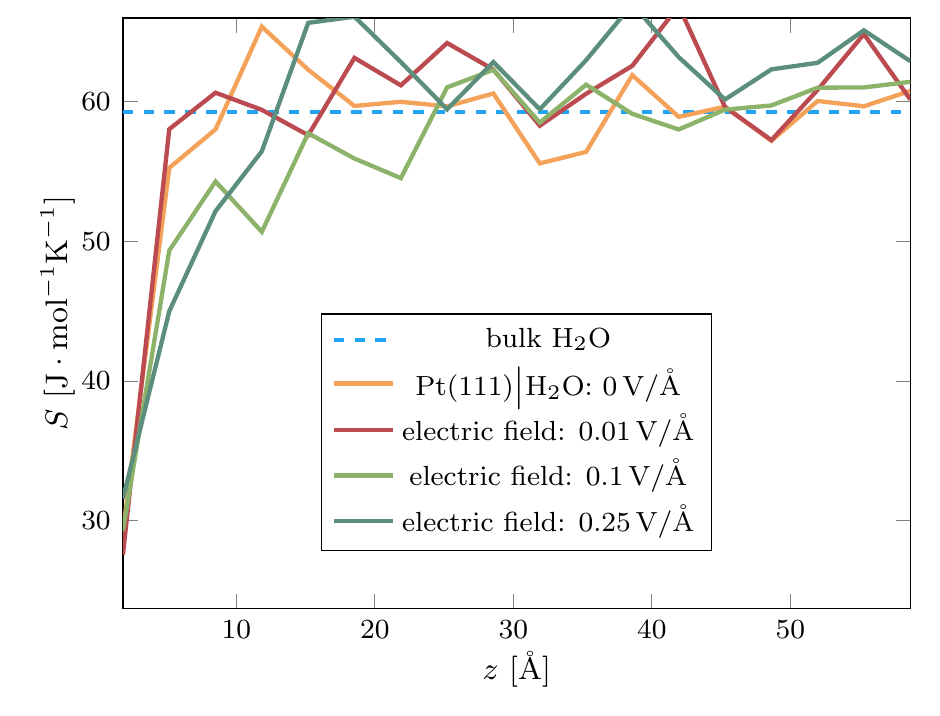}
\caption{Influence of the variation of the external electric field strength on the 2PT calculated entropy $S$. The dashed lines denotes the averaged mean entropy of 59.27\,J/molK for liquid H\textsubscript{2}O.  }
\label{eS}
\end{figure}

As mentioned in the ``ReaxFF methodology'' section, different electric field strengths were applied normal to the surface plane in the Pt(111)\big\vert H\textsubscript{2}O simulations. Here, the influence of an electric field of 0\,V/$\mathrm{\mathring{A}}$, 0.01\,V/$\mathrm{\mathring{A}}$, 0.1\,V/$\mathrm{\mathring{A}}$ and 0.25\,V/$\mathrm{\mathring{A}}$ on the course of the entropy are displayed in figure \ref{eS}. The enhanced ordering character (and lowered entropy near the Pt surface) becomes visible with increasing field strength. While field strengths $<$0.1\,V/$\mathrm{\mathring{A}}$ showed negligible effects on the entropy and the density, stronger electric fields produced diverging entropy and density curves at distances $>$40\,\AA{} from the Pt surface. This can be attributed to the applied thermodynamical ensemble, where the number of H\textsubscript{2}O molecules is kept constant and an accumulation of the water molecules near the Pt(111) surface (\textit{e.g.} increased density) induces a slight depletion further in the bulk.  

\pagebreak

\bibliography{bibtex/LitVerzeichnis1}
\end{document}